\documentclass[epj]{webofc}
\usepackage[varg]{txfonts}   % Web of Conferences font
\newcommand{\be}{\begin{equation}}
\newcommand{\ee}{\end{equation}}
\newcommand{\bea}{\begin{eqnarray}}
\newcommand{\eea}{\end{eqnarray}}

\wocname{EPJ Web of Conferences}
\woctitle{CONF12}
\begin{document}
\selectlanguage{english}
\title{Recent progress in understanding deconfinement\\ and chiral restoration phase transitions}
%
% subtitle (optional, strongly discouraged)

\author{Edward Shuryak \inst{1}\fnsep\thanks
{\email{Edward.Shuryak@stonybrook.edu}} }

\institute{Department of Physics and Astronomy,
Stony Brook University
 Stony Brook NY 11794 USA
}
%\and
%           The second here 
%\and
%           The last address here
%}

\abstract{% 
 Paradigme shift in gauge topology, from instantons to their constituents -- {\em instanton-dyons} -- has recently lead to very significant advances. 
 Like instantons, they have fermionic zero modes, and their collectivization
 at sufficiently high density  explains the chiral symmetry breaking.
 Unlike instantons,
  these objects have  electric and magnetic charges.
  Their back reaction on the mean value of the Polyakov line  (holonomy) 
   allows to explain the deconfinement transition. The talk summarizes recent works
   on the dyon ensemble, done in
   the mean field approximation (MFA), and also by direct numerical statistical simulation.
Introduction of non-trivial quark periodicity conditions leads to drastic changes in both
deconfinement and chiral transitions. In particulaly, in the so called $Z(N_c)-QCD$ model
the former gets much stronger,  while the latter does not seem to occur at all.   
}
\maketitle
\section{Introduction}
\label{intro}

Confinement is the most famous non-perturbative feature of the gauge theories.
Its most intuitive explanation from 1970's is the {\em dual superconductor model} 
by Nambu, 't Hooft and Mandelstam \cite{dual}.  
Lattice studies did reveal the monopoles (albight defined in certain gauges)
and prove that they indeed form Bose-Einstein condensate at $T<T_c$,
see e.g. \cite{D'Alessandro:2010xg}.

Original discussion of the chiral symmetry and its breaking predate QCD and even
quarks. In a classic paper  Nambu and Jona-Lasinio \cite{NJL} showed that an attraction in the scalar $\bar q q$ channel,
if strong enough, can dynamically ``gap" the surface of the Dirac sea. The origin
of this interaction -- claimed to be the origin of the mass of the ``constinuent quarks",
the nucleons and thus ourselves -- was of course in 1961 completely unknown.
Two decades later, in my paper \cite{Shuryak:1981ff},  it was suggested that
NJL attraction is nothing elase but 't Hooft effective $2N_f$ interaction
induced by instantons. Two parameters of the NJL model has been substituted by another two: the
total instanton-antiinstanton density $n\sim 1/fm^4$ and their typical size $\rho \sim 1/3 \, fm$.  
Of course, 't Hooft vertex does more than the NJL operator:  in particular, it breaks the $U(1)_A$ 
symmetry explicitly.

A decade later instantons  were found and studied on the lattice.
Statistical mechanics of instanton ensemble,  including  't Hooft interaction
 to all orders,  known as  
 the Interacting Instanton Liquid Model,  has been developed and solved numerically in 1990's, for a review
  see
\cite{Schafer:1996wv}.  Among other things, it introduced the notion of ``collectivized zero mode zone", or ZMZ for short.  Lattice practitioners
struggle with it till now, since most of numerical fluctuations in simulations come from it.
And yet, some important questions remained unanswered, such as e.g.
{\em Is there any connection between confinement and chiral symmetry breaking?
Why is it that the corresponding finite temperature transitions happen
close, $T_c\approx T_\chi$ for fundamental quarks, but not for adjoint ones \cite{Karsch:1998qj} ? 
Apart of color representations and the number of flavors, can one introduce  other parameters , affecting these transitions
and revealing the underlying mechanism?
}

The so called Polyakov line is used as a deconfinement order parameter, being nonzero at $T>T_c$.
Interpreting this as existence of nonzero average $A_0$ field, one needs to modify all clasical solutions
respectively.  When such solutions were found in 1998 \cite{Kraan:1998sn,Lee:1998bb} it has been realized
that 
instantons  get split into $N_c$ (number of colors) constituents, the selfdual {\em instanton-dyons}\footnote{
They are  called  ``instanton-monopoles" in applications to supersymmetric settings, e.g. 
by Khose et al and Unsal et al. Similar (but no idenical) objects were called 
 the ``instanton quarks" by Zhitnitsky et al.} , 
 connected only by (invisible) Dirac strings.
Since these objects have nonzero electric and magnetic charges and source
Abelian (diagonal) massless gluons, the corresponding ensemble is 
an ``instanton-dyon plasma", with long-range Coulomb-like forces between constituents.  

The first application of the instanton-dyons were made soon after their discovery
in the context of supersymmetric gluodynamics \cite{Davies:1999uw}. This paper solved a puzzling
mismatch (by the factor 2) of the value of the gluino condensate, between the instanton-based and general supersymmetric
evaluations of it. 

Diakonov and collaborators (for review see \cite{Diakonov:2009jq} )
 emphasized that, unlike the (topologically protected) instantons, the dyons interact directly with
 the holonomy field. They suggested that since such dyon (anti-dyon)  become denser
at low temperature, their back reaction  may overcome perturbative holonomy potential and drive it
to its confining value, leading to  vanishing of the mean Polyakov line, or confinement.
Specifically, Diakonov and collaborators focused on the self-dual sector $L,M$ and studied the one-loop
contribution to the partition function \cite{Diakonov:2004jn}. The volume element of the moduli space was
written in terms of dyons coordinates as a determinant of certain matrix $G$, to be referred to as Diakonov determinant. In a dilute limit it leads to
 Coulomb interactions between the dyons, but in the dense region it becomes strongly repulsive, till at certain density
the moduli volume vanishes. 

  A semi-classical  confining regime has been defined by Poppitz et al
~\cite{Poppitz:2011wy,Poppitz:2012sw}   in a carefully devised setting of softly broken supersymmetric models.
 While the setting includes a compactification on a small circle, with  weak coupling and
 an {\em exponentially  small}  density of dyons, the minimum at the confining holonomy
  value is induced by the repulsive interaction in the dyon-antidyon molecules (called  
 $bions$ by these authors). 
 The crucial role of the supersymmetry is the cancellation of the perturbative Gross-Pisarski-Yaffe-Weiss (GPYW)  \cite{Gross:1980br} holonomy potential:
 as a result, in this setting there is no deconfined phase with trivial holonomy at all, unless supersymmetry is softly broken.
  Sulejmanpasic and myself \cite{Shuryak:2013tka} proposed a simple analytic model for the dyon ensemble
with  dyon-antidyon ``repulsive cores", and have shown how they may naturally
 induce confinement in dense enough dyonic ensemble.

Recent progress to be discussed below is related to studies of the instanton-dyon ensembles.
We will focus on a series of papers devoted to high-density phase and mean field
approximation \cite{Liu:2015ufa,Liu:2015jsa,Liu:2016thw,Liu:2016mrk,Liu:2016yij} in section \ref{sec_meanfield},
and on the direct
numerical simulation of the dyon ensembles \cite{Faccioli:2013ja,Larsen:2014yya,Larsen:2015vaa,Larsen:2015tso,Larsen:2016fvs} in section \ref{sec_simulations}. 

 Important ingredient of both of them is classical dyon-antidyon interaction, determined 
 Larsen and myself   in Ref\cite{Larsen:2014yya}: its discussion we skip for space reasons.
We would like to emphasize also importance
of studies devoted to the instanton-dyon identification on the lattice, such as \cite{Bornyakov:2015xao}.

\section{Non-zero holonomy, instanton-dyons and confinement}

\begin{figure}
 \centering
\includegraphics[width=6.cm]{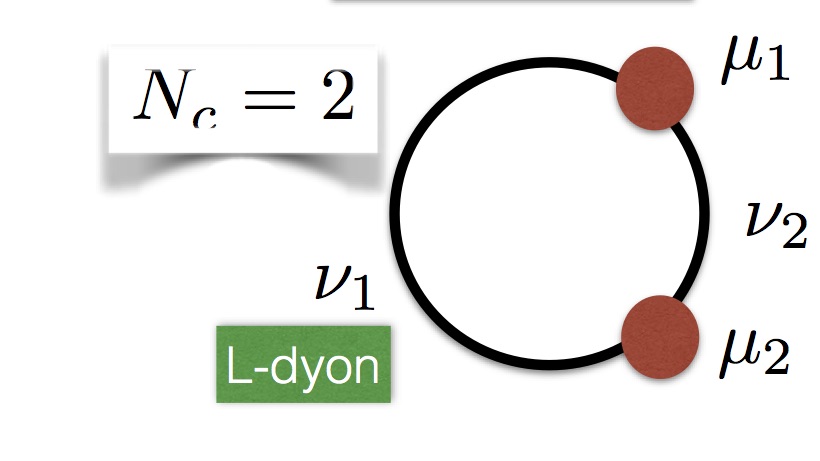}
\includegraphics[width=4.cm]{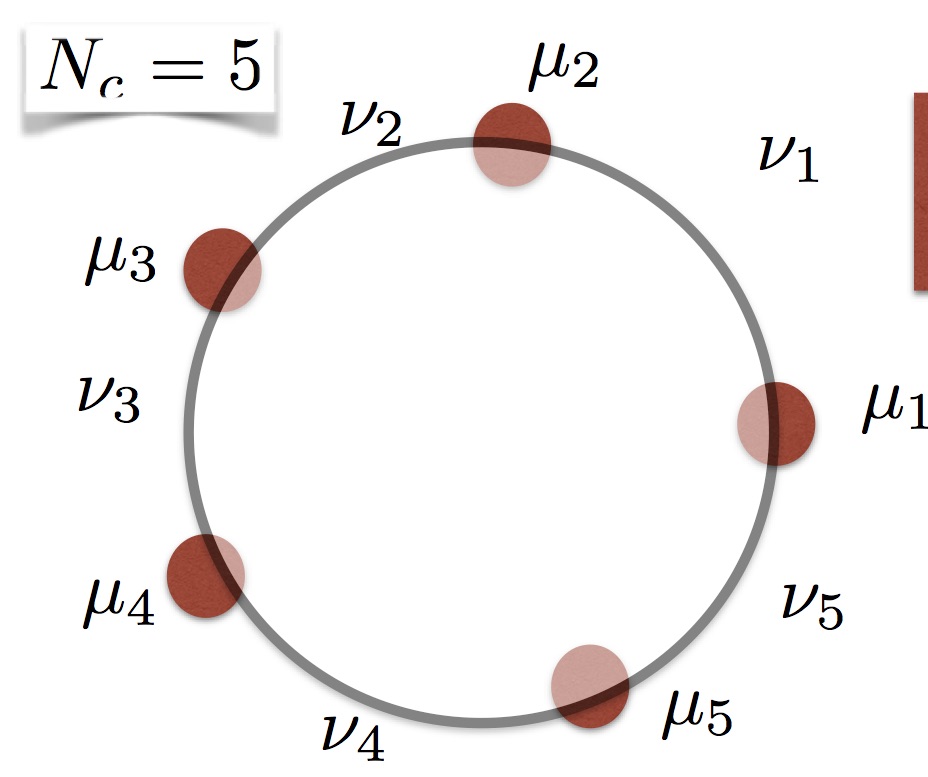}
\caption{ The holonomy circle and the definition of the parameters $\mu_i$ and $\nu_i$, for 2 and 5 colors.}
\label{fig_hol_circle}
\end{figure}

 The ``holonomy" issue refers to 
 the observation that a (gauge invariant) Polyakov loop 
 \be P= Pexp( i\int_0^\beta A^a_0 T^a d\tau ) \ee
has a nonzero vacuum expectation value (VEV), $<P>\neq 0$. Note that there is no trace in this formula, so it is 
some unitary color matrix.
One can always gauge rotate it  to be a diagonal one, to Cartan sub-algebra with $N_c-1$ parameters.
The corresponding 
 phases are denoted $2\pi\mu_i$, and their subsequent differences $\nu_i=\mu_{i+1}-\mu_i$. Fig.\ref{fig_hol_circle}
 explains these notations, for 2 and 5 colors. Note that $\sum \mu_i=0$ and $ \sum \nu_i=1$: thus $\nu_i$ are
 fractions of the holonomy circle. The action and the topological charge of the $N_c$ dyond is $\nu_i$ times that
 of the instanton. %It is in this sense they are ``instanton constituents".   
 
   For the simplest
 $SU(2)$ color group to be discussed below
 there is only one diagonal generator $T^3$ and only one parameter: The notations to be used below are 
$\nu=\nu_2, \bar\nu=1-\nu=\nu_1$.  
 The mean Polyakov line is in this notation simply
 \be < {1\over 2} Tr P>=cos(\pi\nu) \ee
 At high $T$  $<P>\rightarrow  1$, which means all $\mu_i \rightarrow  0$. Thus all but one $\nu_i\rightarrow  0$,
 and one tends to 1: so the original instanton action is recovered.
 In the temperature interval
  $(1..3)T_c$ the mean Polyakov line or $\nu(T)$  is a smooth function of the temperature,
  changing from 0 to 1. Accounting for this phenomenon lead
 Pisarski  and collaborators to  ``semi-QGP"  paradigm \cite{Pisarski:2009zza} and 
 eventually led to the construction of the PNJL model, in which light quark paths are weighted by $<P(T>$. 
  At $T<T_c$, in a confined phase, $<P>=0$ which means that $\nu=1/2$. 
  
A nontrivial average value of the Polyakov line $<P>\neq 1$ , indicating that an expectation value of the gauge potential  
is nonzero . This calls for re-defining the boundary condition of $A_4$ at infinity, for any solitons made of gauge fields,
including the instantons. 

For the SU(2) gauge group
 the selfdual ones are called $M$ with charges $(e,m)=(+,+)$ and $L$ with charges $(e,m)=(-,-)$, the anti-selfdual antidyons are called  
 $\bar{M}$, $(e,m)=(+,-)$ and  $\bar{L}$, $(e,m)=(-,+)$.

%\subsection{Instanton-dyon ensemble and confinement}

\section{Dense dyon plasma in the mean field approximation} \label{sec_meanfield} 

The first paper of the series, by Liu, Zahed and myself \cite{Liu:2015ufa},
had established the mean field  approximation (MFA) in the technical sense. 
The derivation is rather traditional: after bosonization of the partition function, certain fields 
are decleared to be $x$-independent parameters, over which the free energy is minimized.
The derivatives over all parameters define the so called ``gap equations", which needs to be solved
together, defining their values at the global minimum. 
 
 Physics-wise, the main idea
is that if the ensemble of the dyons is dense enough, 
strong electric screening appears, which effectively reduces the
pair-wise correlations in the system, in favor of some average
 mean field. Here there is no  place to
present technical details of these works, and we just summarize the results.

It is shown, that dense enough dyon ensemble does shift the
minimum of the holonomy potential to the confining value, $\nu=1/2$ for the SU(2)
gauge theory considered.  

The next work of the series 
\cite{Liu:2015jsa}
 applies MFA to the $N_c=2$ color theory with $N_f=2$ light quark flavors.
  At high density the minimum of the free energy still corresponds to the confining ensemble with $\nu=1/2$. 
The gap equation for the effective quark mass (or the quark condensate)  of \cite{Liu:2015jsa} the usual form
\be \int {d^3 p \over (2\pi)^3} {M^2(p) \over p^2+M^2(p)}= n_L
\ee
where the r.h.s. is the $L$-dyon density. The equation is for the parameter $\lambda$ in the mass function 
$ M(p)=\lambda p T(p)$, in which $T(p)$ being the Fourier transform of the  ``hopping matrix element"
calculated using the fermionic zero mode. 
Momentum dependence of $M(p)/\lambda$ is shown in Fig.\ref{M_of_p}: note that its shape and the behavior at small momenta is different from that
of an instanton.
 The solution for the condensate can be parameterized as 
\be {| <\bar q q> |\over T^3}\approx 1.25 \left({n_L \over T^3}\right)^{1.63}
\ee

\begin{figure}[htbp]
\begin{center}
\includegraphics[width=8cm]{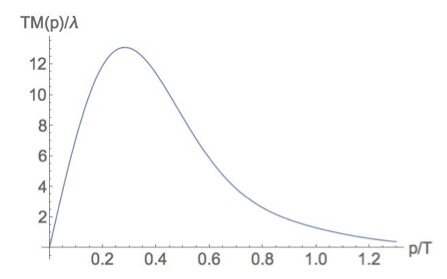}
\caption{The momentum dependent constituent quark mass $T M(p) /\lambda$ versus momentum in units of temperature $p/T$.}
\label{fig_M_of_p}
\end{center}
\end{figure}

A generalization of the mean field treatment to arbitrary number of colors and flavors  in \cite{Liu:2015jsa} shows that 
this gap equation has nonzero solutions for the quark condensate only  if 
\be N_f<2N_c \ee
So,  the critical number of flavors is $N_f=6$ for $N_c=3$.
The lattice simulation indeed show weakening of chiral symmetry violation effects with increasing $N_f$,
but specific results about 
on the end of chiral symmetry breaking are so far rather incomplete: for $N_c=3$ we know that
in the $N_f=4$ case the chiral symmetry is broken, the case $N_f=8$ is not yet completely decided and  $N_f=12$
seems to be already in the conformal window.

Another important generalization -- for quarks in the adjoint representation - is made in a separate paper \cite{Liu:2016mrk}.
The number of fermionic zero modes  increases, and they are more comlicated.
In the symmetric dense phase both M and L dyons have two zero modes. 
But the actual difficulty is not some longer expressions but the fact that one of them has rather 
singular behavior exactly at the confining value of the holonomy, $\nu=1/2$, so approaching it needs special care.
In the case $N_c=2,N_a=1$ the deconfinement and chiral restoration happen at about the same temperature.

\begin{figure}[b!]
  \begin{center}
  \includegraphics[width=9cm]{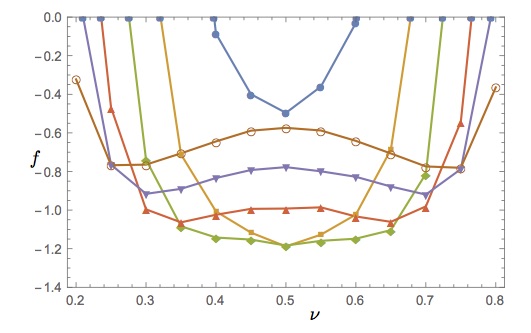}
%  \put(-95,-5){$\nu$}
%    \put(-230,70){$f$}
   \caption{ (Color online). Free Energy density $f$ as a function of holonomy $\nu$ at $S=6$, $M_D=2$ and $N_M=N_L=16$. The different curves corresponds to different densities. $\bullet$ $n=0.53$, $\blacksquare$ $n=0.37$, $\blacklozenge$ $n=0.27$, $\blacktriangle$ $n=0.20$, $\blacktriangledown$ $n=0.15$, $\circ$ $n=0.12$. Not all densities are shown.}
  \label{fig_potential}
  \end{center}
\end{figure} 

\section{Statistical simulations of the instanton-dyon ensembles} \label{sec_simulations}

%\subsection{Instanton-dyon ensemble and chiral symmetry breaking}
The first direct simulation of the instanton-dyon ensemble with dynamical fermions
has been made by Faccioli and myself in \cite{Faccioli:2013ja}. The general setting
follows the example of the ``instanton liquid", it included the determinant'of the so called "hopping matrix", a part of the Dirac operator in the quasizero-mode
sector. It has been done for $SU(2)$ color group and the number of fermions
flavors $N_f=1,2,4$. Except in the last case, chiral symmetry breaking
has been clearly observed, for dense enough dyon ensemble.

 Larsen and myself  \cite{Larsen:2015vaa} use direct numerical simulation of the instanton-dyon ensemble, both in the high-T dilute and low-T dense regime. 
 Unlike the previous work, it uses
 classical dyon-antidyon interaction determined in Ref.\cite{Larsen:2014yya}.
 The holonomy potential as a function
of all parameters of the model is determined and minimized.
 In Fig.\ref{fig_potential} from this work we show the dependence of the 
total free energy on holonomy value, for different ensemble densities.
As one can see, at high density of the dyons their back reaction shifts the minimum to $\nu=1/2$,
which is the confining value for SU(2) ($cos(\pi \nu)=0$):
the confinement transition is thus generated .
 The self-consistent parameters of the
ensemble, minimizing the free energy, is  determined for each density.

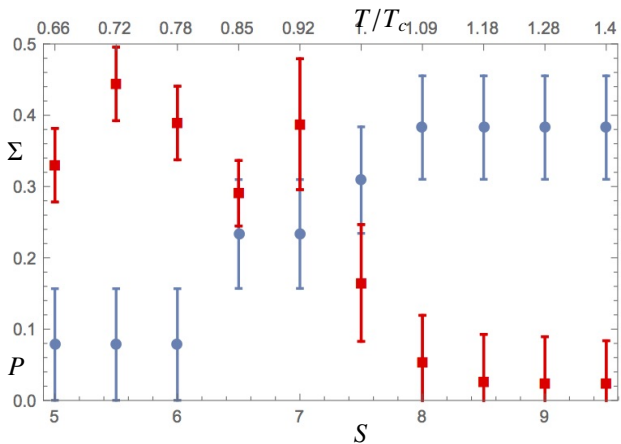
\begin{figure}[h]
\centering
\includegraphics[width=8cm]{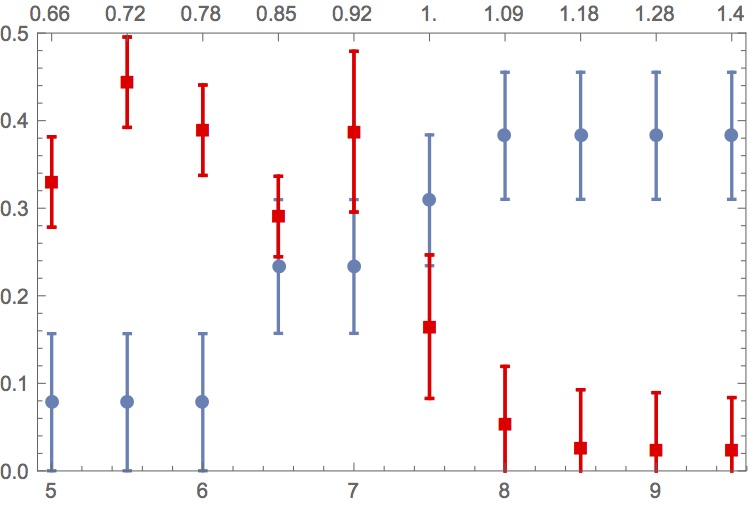}
\put(-100,-5){$S$}
\put(-100,150){$T/T_c$}
\put(-230,100){$\Sigma$}
\put(-230,20){$P$}
\caption{(Color online) The Polyakov loop $P$ (blue circles) and the chiral condensate $\Sigma$ (red squares) as a function of action $S=8\pi^2/g^2$ or temperature $T/T_c$. $\Sigma$ is scaled by 0.2.}
\label{fig_cond}
\end{figure}

The next work of Larsen and myself \cite{Larsen:2015tso} addressed the issue of chiral symmetry breaking 
in the $N_c=2$ theory with two light quark flavors $N_f=2$.
 Numerical simulations are done for partition function appended by the fermionic determinant,
evaluated in the zero mode approximation. 
Using two sizes of the system, with 64 and 128 dyons, we 
identify the finite-size effects in the eigenvalue distribution, and 
extrapolate to infinite size system.
The location of the chiral transition temperature is 
defined both by extrapolation of the quark condensate, from below, and
the so called ``gaps" in the Dirac spectra, from above.
We do indeed observe,  for SU(2) gauge theory with 2 flavors of light fundamental quarks, that  the deconfinement 
and chiral symmetry restoration transitions occur about at the same 
 dyon density, see Fig.\ref{fig_cond}.
 Determination of the precise transition points  
 is difficult 
  since  both transitions appear to be in this case just a smooth crossovers.
  Those should correspond to {\em inflection points} (change of curvature) on the plots
  to be shown. Looking from this perspective at Fig. \ref{fig_cond},
  one would locate the  inflection points of both curves, for  $<P>$ or $<\bar \psi \psi>$ , at
  the same location, namely $S=7-7.5$. 

\section{Quarks with nontrivial periodicity and $Z(N_c)$ QCD}
Periodicity condition along the Matsubara circle can be defined with some arbitrary angles $\psi_f$ for quarks
with the flavor $f$. 
As was determined by van Baal and collaborators, fermionic zero mode ``hops" from one type of dyon to the next
at certain critical values. The resulting rule is: it belogs to the dyon corresponding to the segment of the
holonomy circle $\nu_i$ to which the periodicity phase belongs: $\mu_i<\psi_f<\mu_{i+1}$.

\begin{figure}[h]
\centering
\includegraphics[width=4.cm]{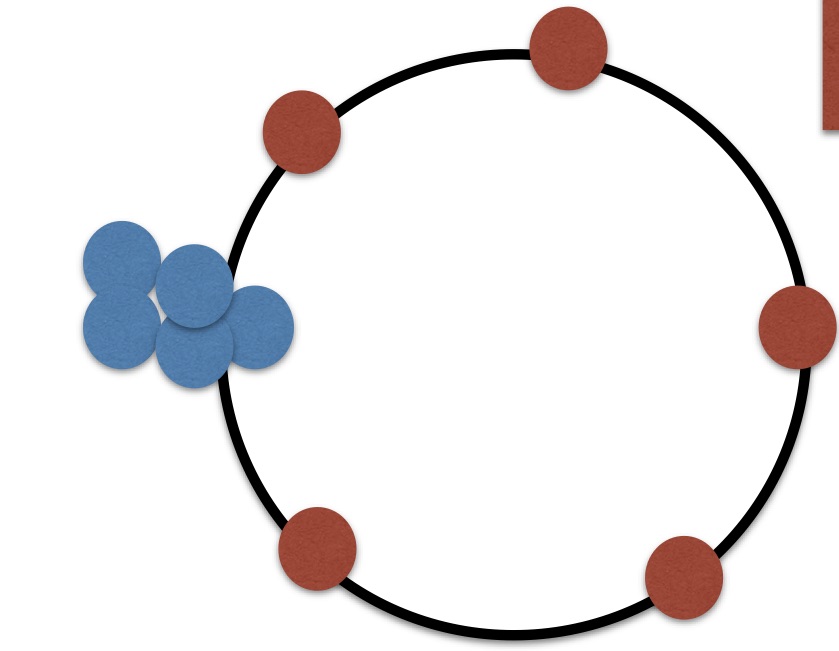}
\includegraphics[width=3.5cm]{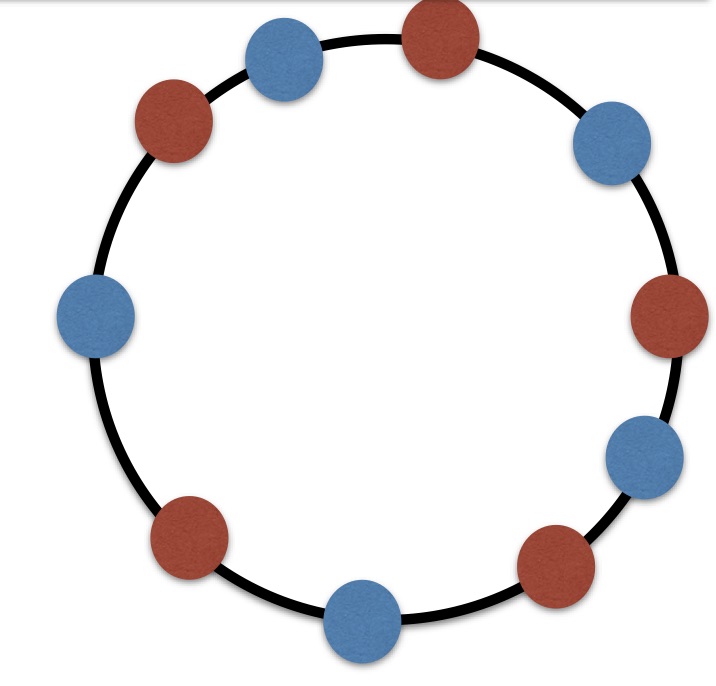}
\caption{ Schematic explanation of the difference between the usual QCD (left) and the $Z(N_c)$ QCD (right).}
\label{fig_circles}
\end{figure}

In physical QCD all quarks are fermions, and therefore $\psi_f=\pi$ for all $f$. This case is
schematically shown by blue dots in Fig.\ref{fig_circles}(left): all fermions fall on the same segment of the
circle, and therefore only one, of $N_c$ dyons, have zero modes and interact with quarks.

But one can introduce other arrangements of these phases. In particular, for $N_c=N_f$ the opposite extreme is 
the so called  $Z(N_c)$ QCD, proposed in \cite{K1,K2,K3,K4,K5} , put them symmetrically around the circle, see Fig.\ref{fig_circles}(right).
In this case, the instanton-dyon framework becomes very symmetric: each dyon interact with ``its own" quark flavor.

The $Z(N_c)$ QCD has been studied  in the mean field framework \cite{Liu:2016yij},  
by statistical simulations \cite{Larsen:2016fvs} and also by lattice simulations \cite{Misumi:2015hfa}.
The first two papers consider the $N_c=N_f=2$ version of the theory, while the last one focus on the
$N_c=N_f=3$. In the former case the set of phases are $\psi_f=0,\pi$, so one quark is a boson and one is a fermion. In the latter  $\psi_f=\pi/3,\pi,-\pi/3$.

\begin{figure}[h]
\centering
\includegraphics[width=7.cm]{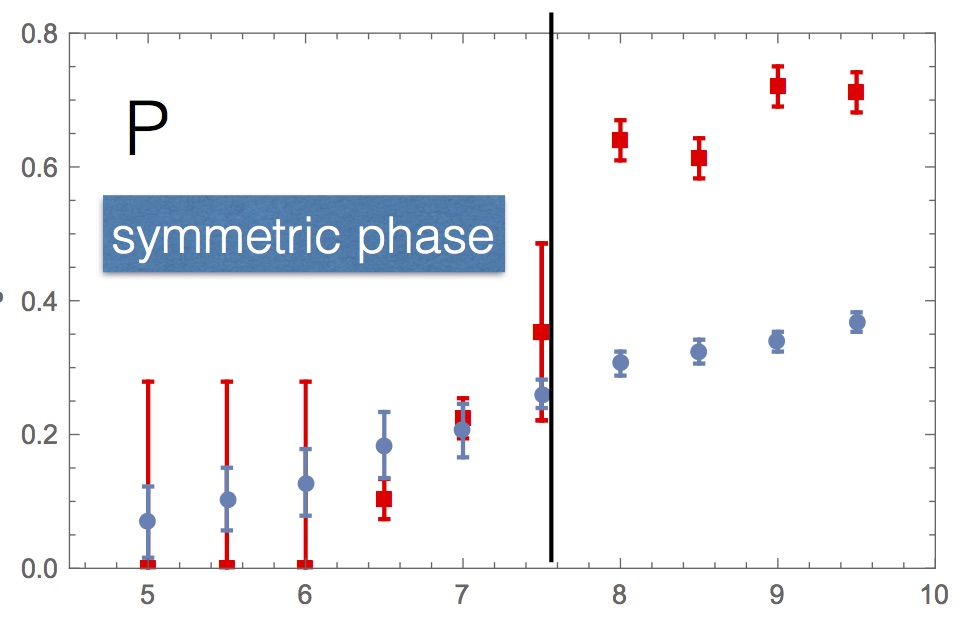}
\includegraphics[width=7.cm]{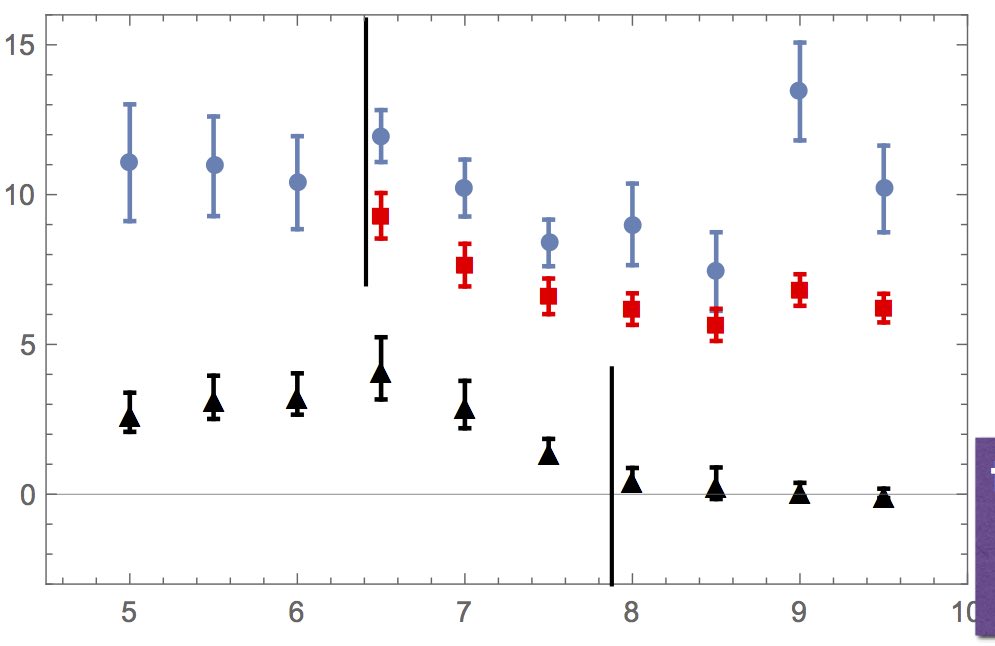}
\caption{(left) The mean Polyakov line P versus the density parameter $S$. Red squares
are for $Z_2 QCD$ while blue circles are for the usual QCD, both with $N_c=N_f=2$ .
(right) The quark condensate versus the density parameter $S$.
Black triangles correspond to the usual QCD: and they display chiral symmetry restoration.
Blue and red poins are for two flavor condensates of the $Z_2 QCD$: to the left of vertical
line there is a ``symmetric phase" in which both types of dyons and condensates are the same.
}
\label{fig_Z2}
\end{figure}

All  these works find deconfinement transition to strengthen significantly, compared to QCD 
with the same $N_c,N_f$ in which it is a very snooth crossover. While in \cite{Liu:2016yij}
the $<P>$ reaches zero smoothly, a la second order transition, the simulations \cite{Larsen:2016fvs}
and lattice \cite{Misumi:2015hfa} both see clear jump in its value indicated strong first order transition.
The red squares at Fig.\ref{fig_Z2}(left)  from \cite{Larsen:2016fvs} are comparing the behavior of the mean Polyakov line
in  $Z_2$ and ordinary QCD. The parameter $S$ used as measure of the dyon density
is the ``instanton action", related with the temperature by
\be S=({11N_c \over 3}- {2N_f \over 3})log({T \over \Lambda})
\ee
The dyons share it as $S_M=\nu S, S_L=\bar \nu S$.
So, larger $S$ at the r.h.s. of the figure correpond to high $T$ and thus to more dilute ensemble,
since densities contain $exp(-S_i)$.

All three studies see a non-zero chiral condensates in the studied region of densities:
perhaps no chiral restoration happens at all. The value fo the condensate are shown in 
 Fig.\ref{fig_Z2}(right)  from \cite{Larsen:2016fvs}.

 The simulation   \cite{Liu:2016yij} demonstrate that
the spectrum of the Dirac eigenvalues has a very specific ``triangular" shape, characteristic
of a single-flavor QCD. This explains why the $Z(N_c)$ QCD has much larger condensate
than ordinary QCD, at the same dyon density, and also why there is no tenedecy to restoration.
As expected, all works see different  condensates, $<\bar u u>\neq <\bar d d>$, but with
difference smaller than one could expect from the difference in the dyon density.

\section{Summary}
  Studies of semiclassical theory and gauge topology shifted from 
  instantons in the QCD vacuum to  the finite temperature
  phenomena, and more specifically, to the mechanisms of the deconfinement and
  chiral restoration transitions.
 Incorporation of nonzero VEV of the Polyakov line -- called holonomy --
 lead to a shift  from instantons to their constituents -- {\em instanton-dyons}.
 Recent papers on the ensembles of those,
 done both in the mean field approximation and by direct statistical simulations, 
  has lead to very significant advances. Unlike instantons,
  these objects have three different set of charges, therefore affecting holonomy value and thus allowing to
  tie   together deconfinement and chiral restoration.

Like the instantons, dyons have topological charges and are subject to
topological index theorems. It means that for appropriate fermionic bondary
conditions they must have fermionic zero modes. Collectivization
of those into a quark condensate follows, provided the ensemble is dense enough.
Unlike instantons, the dyons posess (Euclidean) electric charges, and interact directly
with the holonomy. Therefore, they back-react and are able to modify the holonomy potential. Furthrmore, as the calculations showed, the potential's minimum {\em shifts
to  confining value of the holonomy}, at which all types of dyons become equal. 

Unlike instantons, the dyons posess magnetic charges, and thus their ensemble
generates {\em the magnetic screening mass}. Recall that perturbative polarization
tensor does not generate it \cite{Shuryak:1977ut}: but, according to lattice 
data, in the near-$T_c$ region it even surpasses the electric mass. While we have not discuss it above, let me just mention that it clearly indicates a transition from
electric (QGP) to magnetic plasma, as the coupling grows with decreasing temperature.

This round of studies had shown why QCD with light quarks, unlike pure gauge theories, have rather smooth cross-over
transition: the reason is the symmetry between different dyon kinds is broken by
quarks  in a very robust way: thus  the ``symmetric phase" -- in which all dyons are represented equally
-- is never realized. The densities of $L,M$ dyons are always different by at least factor two or so.

It is a remarkable finding that, with  the same number of light quarks but modified periodicity phases $\psi_f$,
 one gets back a strong first order deconfinement transition.
Further studies with variable phase should reveal an ``internal phase transitions"
related with ``hopping" of quark zero mode from one type of dyon to the next.
Since this phenomenon has no other known explanation, apart of instanton-dyon theory,
documenting those on the lattice would be crucial to finalize the mechanism of the 
deconfinement and chiral transitions.

My final point is an obvious one:
 multiple predictions of the instanton-dyon model
needs to be extensively checked on the lattice. 
 While the periodicity phases mentioned above open new windows,
needless to say, those should also be supplemeneted by a systematic ``hunt" for instanton-dyons
on the lattice, to quantify
 their density and parameters directly.

 \section*{Acknowledgements}
The progress reported
 would not be possible without contributions by Pierre van Baal and
 Mitya Diakonov, the consequences of whose legacy remains to be worked out.
 The  particular results reported here were obtained with my collaborators
 on the instanton-dyon projects, P.Faccioli, T.Sulejmanpasic, R.Larsen, I.Zahed, and Y. Liu.

\end{document}